\documentstyle[aps]{revtex}
\input{epsf}
\begin{document}
\draft
\twocolumn[\hsize\textwidth\columnwidth\hsize
           \csname @twocolumnfalse\endcsname

\title{A system for fast time-resolved measurements of {\it c}-axis quasiparticle conductivity in intrinsic Josephson junctions of Bi$_2$Sr$_2$CaCu$_2$O$_{8+\delta}$}
\author{J.C. Fenton, P.J. Thomas, G. Yang, C.E. Gough}
\address{Superconductivity Research Group, University of Birmingham, Edgbaston, Birmingham B15 2TT, United Kingdom.}
\date{\today}
\maketitle

\begin{abstract}
A wide-band cryogenic amplifier measurement system for
time-resolved four-point IV-characteristic measurements on
Bi$_2$Sr$_2$CaCu$_2$O$_{8+\delta}$ mesa structures is
described. We present measurements which demonstrate the
importance of self-heating on $\sim$ 50 ns time scales.
Such heating is likely to have been very significant in
many previously published measurements, where the reported
nonlinear IV characteristics have been used to derive
superconducting energy gaps.
\end{abstract}
]

\narrowtext

Highly anisotropic high-temperature
superconductors, such as Bi$_2$Sr$_2$CaCu$_2$O$_{8+\delta}$
(2212-BSCCO), can be considered as a linear array of
intrinsic Josephson junctions formed by almost insulating
layers between the superconducting copper-oxide planes.~[1]
To investigate conduction across such junctions, many
groups have published measurements on mesa structures
containing typically 5--20 such junctions.~[2--10]
 The mesas are lithographically patterned [3]
  on the surface of almost ideally atomically flat cleaved surfaces of single crystals of 2212-BSCCO. The aim of such experiments is to investigate intrinsic conduction across the junctions and hence deduce the magnitude and temperature dependence of the assumed {\it d}-wave superconducting gap.

In practice, one observes hysteretic multi-branched
characteristics, corresponding to the expected irreversible
characteristics of each junction in the measured array. In
all previously published measurements, marked deviations
from linearity in the resistive branches have been
reported, as indeed predicted by Josephson tunnelling
theory. In principle, the shape of the characteristics can
be used to determine the superconducting energy gap.~[11]
However, in many circumstances one also observes re-entrant
and even {\it S}-shaped characteristics, which cannot be
described by simple theory. In the past, pulsed
measurements on 200--500 ns time scales have been used in
an attempt to circumvent possible problems from sample
heating.~[6--10]

In this letter, we describe a four-channel cryogenic amplifier
with better than 50 ns resolution (shown schematically in
Fig.~1), which demonstrates the importance of heating even
on sub-$\mu$s time scales.  We report four-point current-voltage (IV)
measurements on a stack of intrinsic Josephson junctions in
a mesa structure of 2212-BSCCO. The mesa structure is
described in more detail elsewhere.~[12] We show that the
characteristics and voltage at which dI/dV becomes infinite
are strongly dependent on the time after the application of
a pulsed current, consistent with heating. This casts
serious doubt on earlier deductions of energy gaps from
such measurements.

Four source-follower buffer amplifiers with high input-impedance
were used.  These were based on the CF739 dual-gate n-channel
depletion-mode field-effect transistor, which has a wide bandwidth and operates
at cryogenic temperatures.  To monitor the output
signals, the amplifier-outputs were connected to a Tektronix
TDS224 100 MHz eight-bit four-channel digital storage oscilloscope via 50
$\Omega$ cryogenic coaxial lines. To retain the large-offset
facility of the oscilloscope without significant loss in
resolution, the amplifiers were designed to have a gain of $\sim$
1/5. The 100 $\Omega$ resistor in series with the input eliminates
high-frequency parasitic oscillations. The response of the
amplifier was linear and almost temperature-independent for inputs
up to 3 V. The small dc offset voltage $\sim$ 0.2 V was
unimportant, since we measured differences in voltage levels
before and during the application of a current pulse. The
amplifiers were fabricated on two circuit boards on either side of
a common earthing plate. These were situated at the
low-temperature end of a continuous-flow cryostat. Short
connections were made to the mesa contacts.

The inputs to the buffer amplifiers were connected to the
mesa structure as illustrated schematically in the insert
of Fig.~2. The voltage difference V$_4-$V$_1$ across the
200 $\Omega$ series resistor monitors the current into the
mesa. The voltage difference V$_2-$V$_3$ corresponds to the
voltage across the shaded region, which for this sample
involved 30 intrinsic junctions.

\begin{figure}[!b]
\epsfxsize=8.5cm
  \epsfbox{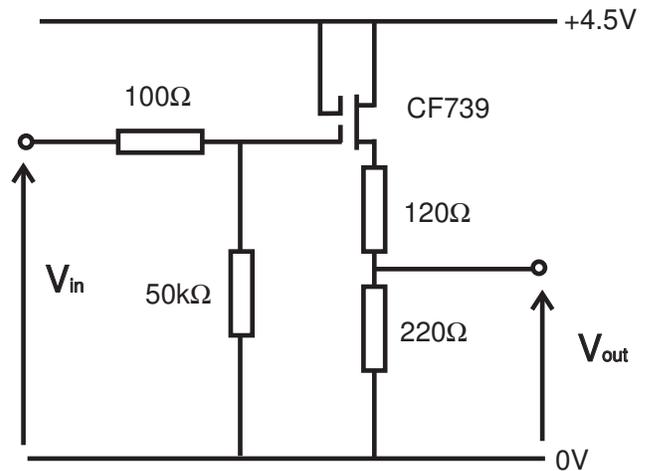}
  \caption{Schematic of circuit for a single buffer amplifier.}
\end{figure}

\begin{figure}[!t]
 \epsfxsize=8.5cm
  \epsfbox{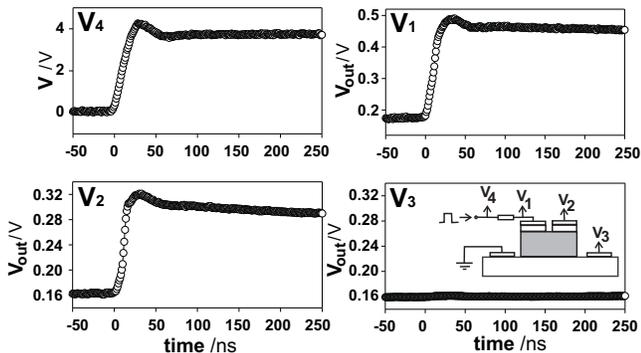}
  \caption{Measured voltages for an applied 4 V pulse at a base
temperature of 10 K.  The insert shows a schematic of
connections to the mesa structure, with four-point
measurements being made across the shaded region.}
\end{figure}
Square-wave current pulses were obtained from a TTi TG1304 50 $\Omega$ output-impedance function generator connected to the sample via a 50 $\Omega$ coaxial line. An initial triangular spike of duration $\sim$ 100 ns could be added at the start of the pulse to briefly exceed the critical currents of the junctions in the stack so that the irreversible part of IV characteristics at low current-bias could be mapped out. However, the data we present here correspond to the characteristics above the critical currents of the individual junctions. In practice, V$_4$ was measured by a 10x probe at the output of the signal generator because, for large inputs, it exceeded the maximum buffer-amplifier
input voltage of 3 V. This introduced a measurable time lag of 1--2 ns between V$_1$ and V$_4$, which we correct for in our analysis of data.

Measurements on short time scales can be affected by
impedance mismatching, leading to unwanted reflections, as
observed by Suzuki {\it et al.}.~[6] We have avoided such
problems by the use of a function generator matched to the
input line and by terminating the coaxial output lines by
50 $\Omega$ connectors at the inputs to the oscilloscope. A
particular advantage of using cryogenic amplifiers is that
the wires between the sample and amplifier-inputs are short
enough for time constants associated with mismatching there
to have a negligible effect on measurements.

Short-time-scale measurements can also be affected by
time constants associated with the resistance and capacitance of
the sample and circuit. For example,
temperature-dependent sample-related rise times $\sim$ 300
ns are evident in several previous measurements.~[9,10] We
have been able to achieve circuit rise times less than
$\sim$ 10 ns, significantly less than the rise time of the
pulse generator.

Typical measured voltages are shown in Fig.~2. The finite
rise time and slight overshoot in the voltage reflect the
``square'' pulse shape from the signal generator and are
almost identical to those measured across a 50 $\Omega$
resistor connected directly to the output of the signal
generator. Any changes in the initial rise time and pulse
shape can therefore be associated with sample properties.

\begin{figure}[!t]
 \epsfxsize=8.5cm
  \epsfbox{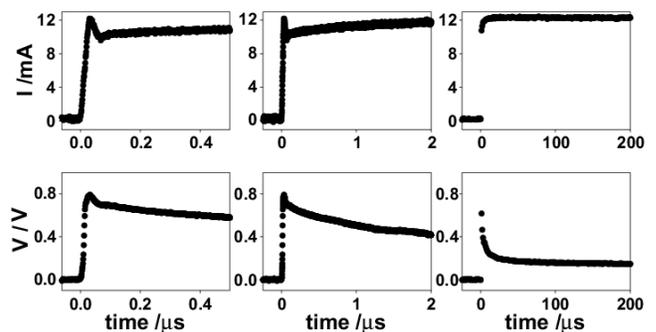}
  \caption{Variation of sample current and voltage at 9 K for an
applied simple square pulse of 4 V.}
\end{figure}
In Fig.~3, we show the current through the sample and the
voltage across the sample on three different time scales
after the application of a square current pulse. We note
the very significant (10\%) drop in voltage after only 500
ns. We identify the drop in voltage with sample heating and
the associated increase in conductivity with temperature.
This increase is shown in the inset of Fig.~4, obtained
from the linear term in fits to a functional form I =
AV+BV$^3$ for measurements at very short times (50 ns).
These measurements are in excellent agreement with earlier
dc measurements of Thomas {\it et al.}~[12] on the same
sample, in which the mesa temperature was measured
directly, allowing corrections for heating in the measured
IV characteristics to be evaluated. We believe the 50 ns
characteristics to be a close representation of the
intrinsic sample properties.

Figure 4 illustrates a series of IV measurements at a
sample base-temperature of 33 K, taken at various times
after the switching-on of square current pulses of
different
\begin{figure}[!b]
\epsfxsize=8.5cm
  \epsfbox{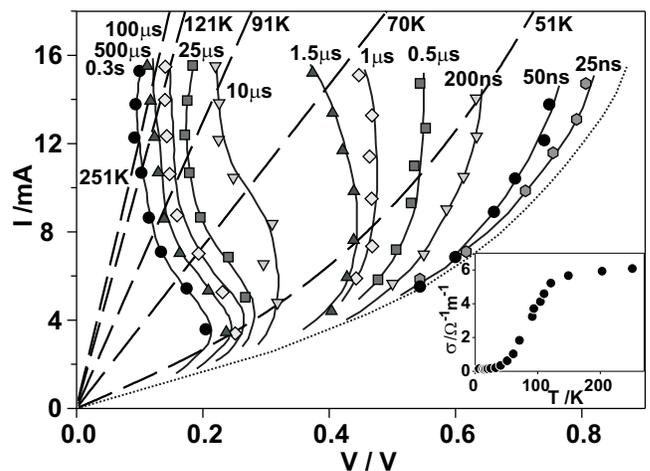}
 \caption{Evolution of IV characteristics with time at a base
temperature of 33 K. Dashed lines show fits to 50 ns IV
characteristics at a number of base temperatures. Solid
lines are guides for the eye. The form of the intrinsic
characteristics at 33 K is suggested by the dotted line.
The inset shows the temperature-variation of the
low-voltage conductivity, determined by a fit of the
early-time data to I = AV+BV$^3$ at each temperature.}
\end{figure}
amplitudes. At short times, the IV characteristics show
little curvature but become increasingly nonlinear at
higher bias.  At the highest bias levels currently
achievable, the slope approaches infinity only 200 ns after
the start of the pulse and significant backbending is
observed 1.5 $\mu$s after the start of the pulse. {\it
S}-shaped IV characteristics are observed 25 $\mu$s after
the start of the pulse, and the 0.3 s characteristics are
indistinguishable from dc measurements. The strong
time-dependence of the shape of the characteristics makes
it clear that the infinite slope and onset of the
backbending feature is not an intrinsic feature associated
with the gap voltage at 33 K.

We deduce the temperature at points on the heated curves by comparison with the 50 ns quasi-intrinsic characteristics at higher base temperatures. For example, for a current of 15 mA, the sample temperature rises from 33 K to above T$_c$ in around 10 $\mu$s, and to above room temperature in about 500 $\mu$s.

Note also that the backbending feature does not correspond to the sample being heated through T$_c$. Neither does it correspond to any feature that might be associated with the gap at the higher temperature to which the mesa is heated. This is illustrated by the onset of infinite slope at around 0.5 V in the 500 ns data, intersecting with the relatively featureless quasi-intrinsic 51 K characteristics. At this temperature, any marked feature associated with the gap parameter clearly lies at larger voltages and larger current bias than those currently available in our measurements. Other than immediately following the application of a pulse, any identification of the energy gap with the voltage at which dI/dV becomes infinite is therefore very unsafe; such values will seriously underestimate the true gap when heating is significant.

Backbending is a consequence of significant heating and the
steep increase in electrical conductivity with temperature.
The {\it S}-shaped curves arise because, once the mesa is heated
to a high temperature, any further rise in temperature with
increasing current has only a small influence on the
electrical conductivity (see inset, Fig.~4).

The problem of self-heating in the steady state in such
measurements has been modelled by Krasnov {\it et
al.}.~[13] Indeed, when a temperature-independent thermal
diffusivity D is assumed and dissipation in the contacts is
neglected, it is straightforward to show that, for a mesa
structure of area L$^2$ containing N intrinsic junctions,
the temperature of the mesa will rise above the base
temperature by an amount $\propto$ NL for a given voltage
across individual junctions in the stack. Heating can be
significantly reduced by using thinner mesa structures
containing a smaller number of junctions, though potential
problems of heating from contact resistances remain.  For
our sample, the contact resistance is comparable to that of
the mesa stack.

Smaller-area junctions can also be used to reduce heating.
However, the initial rate of rise in temperature is
essentially a one-dimensional problem, since the volume of the crystal
initially heated by the mesa is $\sim $
L$^2\sqrt{\textrm{Dt}}$. This gives a temperature rise
which is proportional to t$^{1/2}$ and to the input power
density, but which is {\it independent} of the mesa area
for a given voltage across individual junctions. Therefore,
on time scales t $\ll $ L$^2$/D, self-heating is not
reduced by using smaller-area mesas. Using published data
for the thermal conductivity and heat capacity, we estimate
that at 4 K, L$^2$/D $\sim$ 1 $\mu$s, while at 40 K,
L$^2$/D $\sim$ 500 $\mu$s.~[14,15]

Our measurements on (30 $\mu$m)$^2$ devices already
demonstrate the need to make measurements on $<$ 100 ns
time scales, if one wishes to measure the intrinsic
characteristics at higher bias levels. It is interesting to
note that, in Ref.~[6], Suzuki {\it et al.} made their
measurements on similar-size samples to our own. They
derived characteristics from measurements made 500 ns after
the application of a current pulse, using electronics with
a settling time of $\sim$ 300 ns. However, if they had
derived characteristics from measurements at 800 ns (see
Fig.~5 in Ref.~[6]), a backbending feature at around 0.5 V
would have been observed. This suggests that heating may
well also have been significant in their 500 ns
measurements. [16]

  In summary, a system has been developed to allow measurements to be made on BSCCO multi-junction array mesa structures within 50 ns of the start of the current pulse. This enables us to monitor and largely circumvent problems of self-heating, and thereby to estimate the intrinsic superconducting tunnelling properties between the atomic planes, even when dc measurements would heat the sample to temperatures well above T$_c$. Our measurements demonstrate that backbending and {\it S}-shaped features can be explained by heating without invoking nonequilibrium effects.~[17,18] We suggest that, in most previously reported measurements, the voltage at which backbending was first observed was almost certainly associated with self-heating and not the gap voltage.

This work was supported by the UK EPSRC, grant number GR/L65581.

\end{document}